\input harvmac
\noblackbox
\def\ias{\vbox{\sl\centerline{Department of Physics, University of 
California at San Diego}
\centerline{9500 Gilman Drive, La Jolla, CA 92093 USA}}}

\lref\one{K. Intriligator, Nucl. Phys. {\bf{B496}} (1997) 177,
hep-th/9702038.}
\lref\two{M. Douglas and G. Moore, hep-th/9603167.}
\lref\three{P. Aspinwall, Nucl. Phys. {\bf{B496}} (1997) 149, hep-th/9612108.}
\lref\four{J. Blum and K. Intriligator, Nucl. Phys. {\bf{B506}} (1997) 199,
hep-th/9705030.}
\lref\five{J. Blum and K. Intriligator, Nucl. Phys. {\bf{B506}} (1997) 223,
hep-th/9705044.}
\lref\six{N. Seiberg and E. Witten, Nucl. Phys. {\bf{B471}} (1996) 121,
hep-th/9603003.}
\lref\witten{E. Witten, hep-th/9507121.}
\lref\eight{A. Hananay and Ori Ganor, Nucl. Phys. {\bf{B474}} (1996) 122,
hep-th/9602120.}
\lref\nine{P. Aspinwall and D. Morrison, Nucl. Phys. {\bf{B503}} (1997) 533,
hep-th/9705104.}
\lref\ten{K. Intriligator, hep-th/9708117.}
\lref\eleven{I. Brunner and A. Karch, hep-th/9712143.}
\lref\twelve{A. Hananay and Alberto Zaffaroni, hep-th/9712145.}
\lref\thirteen{E. Witten, Nucl. Phys. {\bf{B460}} (1996) 541, hep-th/9511030.}
\lref\fourteen{J. D. Blum, Nucl. Phys. {\bf{B507}} (1997) 245, hep-th/9702084.}
\lref\fifteen{C. Vafa, Nucl. Phys. {\bf{B273}} (1986) 592.}
\lref\hetone{A. Dabholkar, Phys. Lett.  {\bf{B357}} (1995) 307, 
hep-th/9506160\semi
C. Hull, Phys. Lett.  {\bf{B357}} (1995) 545, hep-th/9506164\semi
J. Polchinski and E. Witten, Nucl. Phys. {\bf{B460}} (1996) 525, 
hep-th/9510169.}
\lref\douglas{A. Connes, M. Douglas, and A. Schwarz, hep-th/9711162.}
\lref\seiberg{O. Aharony, M. Berkooz, and N. Seiberg, hep-th/9712117.}
\lref\oova{H. Ooguri and C. Vafa, Nucl. Phys. {\bf{B463}} (1996) 55, 
hep-th/9511164.}
\lref\seibergtwo{N. Seiberg, Phys. Lett.  {\bf{B388}} (1996) 753, 
hep-th/9608111.}
\lref\gabzw{M. Gaberdiel and B. Zwiebach, hep-th/9709013.}
\lref\pol{J. Polchinski, S. Chaudhuri, and C. Johnson, hep-th/9602052\semi
J. Polchinski, hep-th/9611050.}

\Title{\vbox{\baselineskip12pt
\hbox{UCSD/PTH 97-40}\hbox{hep-th/9712233}}}
{\vbox{\centerline{NS Branes in Type I Theory}}}

{\bigskip
\centerline{Julie D. Blum}
\bigskip
\ias

\bigskip
\medskip
\centerline{\bf Abstract}

We consider novel nonperturbative effects of type I theories compactified
on singular ALE spaces obtained by adding NS branes.  Such effects include
a description of small $E_8$ instantons at singularities.}

\Date{12/97}

\newsec{Introduction}

This brief note proposes that noncompact type I orbifolds can be modified
by the addition of NS branes.  Our focus here will be on orbifold 
compactifications of type I theory to six dimensions with $N=1$ supersymmetry.
In Ref.~\one\ spacetime anomaly considerations were used to
constrain the form of low energy six-dimensional theories resulting from 
small $Spin(32)/{\bf Z_2}$ (heterotic) instantons at 
${\bf C^2/Z_N}$ singularities \two\
and to conjecture phase transitions involving extra tensors.  The N=2 case
had previously been worked out in F theory \three .  In Ref.~\four\ exactly
these theories were derived in string theory as type I orbifolds using
orientifold consistency conditions as well as tadpole constraints.  Related
theories (including $Spin(32)/{\bf Z_2}$ instantons on nonabelian ALE spaces) 
were derived in Refs.~\five\nine .  The spacetime anomalies were derived 
in a simple way from the tadpoles.  That these theories could make sense
at higher energies could be seen from the tadpoles through such effects as
the anomaly inflow mechanism.  It is satisfying that perturbative string
theory gives a microscopic description of such spacetime phenomena.

We will also demonstrate that small $E_8$ instanton (tensionless
string) theories \six\eight\ on noncompact spaces can be understood
via orientifolds of type IIB discussed in \one\four .  One of the subtleties
of the string theory construction allows for NS branes to be added to type I
theories at ALE singularities to obtain, for example, the theories of 
small $E_8$ instantons at $\bf{Z_N}$ singularities found by \nine\ and 
discussed further by \ten .  Before I was able to obtain a microscopic (string
theory) derivation of the twisted charge cancellation condition for the 
theories with NS branes, two papers \eleven\twelve\  appeared which 
engineered some of these theories with branes.

\newsec{Theories with NS branes}

For type IIB theory on a smooth manifold, the operation $\Omega$ or 
worldsheet parity projects out the NS-NS antisymmetric tensor while
retaining the R-R antisymmetric tensor.  The S-duality symmetry of IIB which
exchanges these two tensors as well as inverting the coupling is not
manifest in the perturbative orientifold.  We are allowed to add D5 branes
but not NS5 branes to the perturbative orientifold.  Fundamental type I
strings and D strings are very different objects perturbatively.  We know,
however, that S-duality interchanges type I and the $SO(32)$ heterotic 
theories.  We
also know that the zero size heterotic instanton is a heterotic NS brane
\witten\ and that the D string of type I is equivalent to the fundamental
heterotic string \hetone .  Thus, in some sense, the S-duality subgroup of
the type IIB $SL(2,{\bf Z})$ symmetry is unbroken by the orientifold.
These are the only vacuums that we know how to describe in perturbative
string theory, but could there be other supersymmetric vacuums related
by IIB $SL(2,{\bf Z})$ perhaps describable in matrix theory similarly
to \douglas ?

In \four\five\ it was found that a nonstandard projection of worldsheet parity
($\Omega$) was required for some twisted sectors of the ALE orientifold.
In such sectors the NS-NS antisymmetric tensor is not projected out, but the
R-R antisymmetric tensor is projected out.  In the correspondence between the 
discrete subgroups of $SU(2)$ and the extended Dynkin diagrams of the $ADE$
groups, the nodes of the diagram represent conjugacy classes or twisted 
sectors.  The unusual projection occurred in self-conjugate sectors other
than the untwisted sector (for cases with vector structure) and in half of
the sectors that were conjugate to another sector.
Associated with the twisted
sector where the unusual projection occurs
 is a two-cycle on which a kind of discrete torsion \fifteen\ acts reversing
the usual projection.  Because of this fact, the cycle can act as a source
for the NS-NS field strength--we can add NS branes to the orientifold.

How should we describe such theories perturbatively?  In the general case
these theories are likely to be quite complicated since there will be both
D and NS branes.  Perturbative string theory may be inadequate here, and
considerations of \douglas\seiberg\ are perhaps applicable.  There is,
however, a special case where a string theory description may be viable.
Some of the $A$ theories discussed in \one\four\ have no enhanced gauge
symmetry or D branes.  This is not true for the $D$ and $E$ cases on the
Coulomb branch \five .  In such a case, it may be possible to relate
vacuums with extra NS branes to ones with D branes by S-duality.  I
have not succeeded in finding the precise relation.
In \one\four\five\ the theories of D branes at ALE singularities were
described by extended Dynkin diagrams.  Since the NS-NS charge is always
projected out in the untwisted sector, we expect that these theories
should be described by regular Dynkin diagrams.  We will give an example in 
our discussion of small $E_8$ instantons at ${\bf Z_N}$ singularities.
One would expect that worldsheet parity is exchanged with left-handed
fermion number ($(-1)^{F_L}$).
Since $(-1)^{F_L}$ projects out either the vector or its supersymmetric 
partner, supersymmetry requires that NS branes only be associated with
cycles that are related by complex conjugation to another cycle.  Thus, the
$\rm{N^{th}}$ cycle for the ${\bf Z_{2N}}$ case has no NS branes.  
It is important to note that these theories
with NS branes only make sense in the noncompact case where the total 
number of NS branes is unconstrained.  The requirement to cancel twisted 
NS-NS charge possibly allows for more general configurations of NS branes 
that violate supersymmetry.

\newsec{Small $E_8$ Theories}

The theories to be discussed here are those of $SO(32)$ heterotic instantons
on a noncompact orbifold ${\bf C^2/Z_{2N}}$ with a flat connection at infinity
breaking $SO(32)$ to $SO(16)\times SO(16)$.  When gravity is decoupled, the
resulting theories are $N=1$ gauge theories in six dimensions.  The theory of
the instanton when it shrinks to zero size \thirteen\ can be understood in
type I perturbation theory.  The ninebrane gauge group $SO(32)$ is broken
to $SO(16)\times SO(16)$ by a ${\bf Z_2}\in {\bf Z_{2N}}$ action on the 
Chan-Paton factors.  A Wilson line in the vector representation of $SO(32)$
has $16$ $1$'s and $16$ $-1$'s.
Small $SO(32)$ instantons at the singularity are
described by fivebranes.  Thus, one performs a  ${\bf Z_2}\times {\bf Z_{2N}}$
orientifold of type IIB where the $\bf{Z_2}$ is worldsheet parity.
Consistency conditions and tadpole anomalies constrain the small instanton 
gauge group to be
\eqn\gauge{Sp(K)_1\times U(2K)^{N-1}\times Sp(K)_2}
with hypermultiplets ${1\over 2}(16,2K)_1\oplus {1\over 2}(16,2K)_2$ and 
bifundamentals in adjoining gauge groups.  The Fayet-Iliopoulos mechanism
reduces the $U(K)$'s to $SU(K)$'s.  One also obtains $N$ extra tensor
multiplets and $N-1$ hypermultiplets from twisted sectors.  The $N=1$
supergravity as well as an extra tensor multiplet and some extra
hypermultiplets are obtained from the untwisted sector.  Irreducible gauge
anomalies vanish in these theories, and other anomalies are cancelled through
the Green-Schwarz and anomaly inflow mechanisms.

If we consider the case $K=0$, index theorems (see \five\ ) say that the
instanton number equals $2N$ which is the Euler characteristic of the orbifold.
There is then no enhanced gauge symmetry, and no fivebranes can travel away
from the singularity.  The Coulomb branch with extra tensors 
looks the same as that
of $N$ small $E_8$ instantons described by M theory fivebranes on $S^1/{\bf
Z_2}\times R^4$ with one missing hypermultiplet that can be interpreted as
an overall center of mass related to fixing the position of the ALE 
singularity.  As discussed in \ten\ , the $N$ tensor multiplets live in 
the $Sp(N)$ Coxeter box just like the small $E_8$ instantons.
 The dimension of the Higgs branch describing large instantons and a
blown up orbifold is $30N-1$.  If we consider these instantons to be $SO$
instantons, the classical theory on the whole ALE space makes no sense for
zero size instantons (leads to anomalies).  Thus, we conjecture that these
standardly embedded theories actually describe $N$ $E_8$ instantons with
a missing center of mass mode and scale sizes related to blow up modes.
Another argument is the following.  In \fourteen\ I described sixbranes
compactified on an orientifold of type IIA $T^5\times S^1/{\bf Z_2}$.  By
adding $N$ type IIA NS branes and decompactifying $T^5$ we can obtain the
theories \gauge .  If we now remove the sixbranes and do an S-duality, 
(interchange $S^1/{\bf Z_2}$ and the M theory $S^1$) we
obtain the theory of $N$ small $E_8$ instantons.  I have not worked out in
detail the subtleties of a T-duality map of these two configurations.

Using our idea for adding NS branes to type I orbifolds, we conjecture 
that small $E_8$ instantons at ${\bf Z_k}$ singularities can be described by
adding NS branes to the above description of small $E_8$ instantons.  As
conjectured by \witten\ for ${\bf Z_2}$ and shown for general ${\bf Z_k}$
by \oova , type II theories on $A_n$ ALE spaces transform under T-duality
into $n+1$ NS branes.  That is the motivation for this conjecture.  As shown
by \nine\ the enhanced gauge symmetry obtained from $N$ small $E_8$ instantons
at a ${\bf Z_k}$ singularity is
\eqn\inst{\prod U(v_i)}
where $i$ runs from $1$ to $N$, $v_i=i$ for $i<k$, $v_i=k$ for $k\leq i\leq
N-k$, and $v_i=N-i$ for $N-k< i\leq N$.  The matter includes bifundamentals 
and an extra fundamental of $U(v_k)$ and $U(v_{N-k})$ as shown by \ten .
Again the Fayet-Iliopoulos mechanism reduces $U(v_i)$ to $SU(v_i)$.  As we
have discussed in the previous section, we expect that the theories with
NS branes (no D branes) should be described by regular Dynkin diagrams.
In fact the equation
describing $N$ small $E_8$ instantons at a $\bf{Z_k}$ singularity is
the following.
\eqn\cartan{C_{ij}v_j=\delta_{i,k}+\delta_{i,N-k}+\delta_{i,2N-k}+
\delta_{i,N+k}-2\delta_{i,N}}
Here, $C_{ij}$ is the Cartan matrix of $SU(2N)$ and $v_j$ is the number of
NS branes on the $j^{th}$ cycle, and this equation describes the gauge theory 
in the usual way.  

In \twelve\ it was noted that these particular theories agree with ones that 
were obtained in \one\four .  These theories are described in section $4.1$
of \four .  In the notation used there $K=0$, $P=N$, $W_0=W_N=7$, and
$W_k=W_{N-k}=W_{N+k}=W_{2N-k}=1$.  For $K=0$ this theory is also described
by the equation \cartan\ and is a good candidate for the S-dual of the
theory with NS branes.  As in the above discussion, considering this theory
to describe $SO$ instantons leads to anomalies everywhere on the ALE space
for zero size instantons so that arguments of \pol\seibergtwo\ suggest
that the symmetry is enhanced to $E_8\times E_8$.  The precise action of
S-duality is far from obvious to me.

As we previously mentioned adding NS branes to the $D$ and $E$ singularities is
not likely to have a perturbative string description because these
singularities always have D branes on the Coulomb branch.  
We can also consider the above $E_8$ 
theories with D branes in addition to NS branes.  As shown in \gabzw\ one
can obtain exceptional groups from this configuration through multipronged
string junctions.  This would perhaps be a starting point for reproducing
the $D$ and $E$ theories of \nine\ in string theory.

\bigskip
I thank K. Intriligator
for useful discussions and encouragement.
I was partially supported by NSF grant PHY-9513835.

\listrefs
\end